\documentclass{article}
\usepackage{times}
\usepackage{amsmath}
\usepackage{amsfonts}
\usepackage{amssymb}
\usepackage{latexsym}
\usepackage{graphicx}
\usepackage{bm}
\usepackage{verbatim}

\begin{document}

\begin{center}
\Large{Hydrodynamics of Interacting Rods}\\
\vspace*{2ex}
George D. J. Phillies\\
Department of Physics\\
Worcester Polytechnic Institute\\
Worcester, MA 01609\\
phillies@4liberty.net\\
\date{\today}
\end{center}

\centerline{\bf Abstract}
\vspace*{2ex}
This note considers how hydrodynamic interactions between a pair of rods slow the motion of each rod.  Rods are approximated as being nearly straight lines of pointlike beads. Hydrodynamic interactions between the beads are treated with generalizations, developed by Kynch (\emph{J. Fluid Mech.} \textbf{5}, 193-208 (1959)) and by Mazur and van Saarloos (\emph{Physica A} {\bf 115}, 21-57 (1982)) of the Oseen and Rotne-Prager hydrodynamic interaction tensors. We find that the hydrodynamic interactions between two rods, as they affect the mobility of a given rod, are with a minor exception not very different from the interactions between two beads, especially when the two rods are even modestly separated. At the level of approximation of bead-spring models of polymer coils, it therefore does not appear to be advantageous to replace sphere-sphere hydrodynamic interaction tensors with more elaborate rod-rod hydrodynamic interaction tensors.

\section{Introduction}

This note considers an aspect of a model for polymer dynamics in solution.  Because polymer chains have persistence lengths, one might approximate a polymer as a series of short elastic segments, whose motions except for their considerable elasticity would resemble the motions of the bars of a bar anchor chain. The issue treated here is the nature of hydrodynamic interactions between these segments.

Hydrodynamic interactions appear already in calculations of transport in solutions of colloidal spheres. At large distances, these interactions are described by the Oseen tensor.  The Oseen tensor describes the part of the sphere-sphere hydrodynamic interaction that decreases inversely as the first power of the distance between the spheres.  The Oseen tensor is inaccurate at short ranges, where hydrodynamic interactions between spheres are expected to be the strongest.  As shown by Kynch\cite{kynch1958a} and by Mazur and van Saarloos\cite{mazur1982a}, among others, sphere-sphere hydrodynamic interaction tensors can be expanded in power series in (negative) powers of the distances $r$ between pairs, trios, quartets, etc., of interacting spheres.  The effects of short-range sphere-sphere hydrodynamic interactions on the self- and mutual-diffusion coefficients of colloidal spheres have been calculated\cite{carter1985a,phillies2016a}; the effects are significant.

Here we are interested in the motions of rigid assemblies of spheres. Close to four decades ago, there was substantial scientific interest in calculating the hydrodynamic transport properties of rigid assemblies of small spheres, as a path to modelling the hydrodynamic properties of objects of irregular shape\cite{bloomfield1967a,bloomfield1967b,torre1978a,bernal1981a,allison1981a}. Hydrodynamic interactions between the spheres in an assembly change the flow field around the assembly, so the hydrodynamic drag coefficient of an assembly of spheres of some shape is not simply the sum of the drag coefficients of the individual spheres.  When this approach was used to compute the drag coefficient of simple objects, such as ellipsoids, whose drag coefficients were known from analytic calculations, the modeled approach agreed with exact results to within a few percent. Most of these calculations invoked the Oseen or Rotne-Prager forms for sphere-sphere interactions.

This author\cite{phillies1984z} previously examined the same problem, taking into account the higher-order hydrodynamic interaction terms obtained by Kynch and by Mazur and van Saarloos.  These higher-order terms were complete through order $r^{-7}$. For the sphere arrangements that were studied, higher-order hydrodynamic interactions tended to cancel each other, so models based on the Oseen and Rotne-Prager tensors were found to be more accurate than might have been anticipated.

Modern models of the dynamics of polymers in solution were early elaborated by Kirkwood and Riseman\cite{kirkwood1948a}.  These authors describe a polymer coil as a statistical bag of beads, distances between pairs of beads being taken to be their average values as determined by their separations along the polymer chain and the presumed Gaussian random-walk form of a polymer coil in solution.  As shown by Kirkwood and Riseman, in the presence of a velocity shear field in the surrounding solvent, a polymer coil translates and rotates.  Individual beads in the coil cannot at every point move at the local velocity of the solvent, so viscous dissipation occurs. The local solvent velocity was taken by Kirkwood and Riseman to be the sum of the velocity field due to bulk solvent motion, and the fluid flow created at each point in the system by the motions of the polymer beads with respect to the solvent. In the original calculations, fluid flow arising from bead motions was taken to be adequately described by the Oseen hydrodynamic tensor.

A Kirkwood-Riseman polymer coil has only two modes of motion, namely (i) uniform translation and (ii) simple rotation, rotation like the rotation of a rigid body. These modes are vectors, so there are six independent components to them. In either of these modes, the positions of the beads relative to each other do not change. Kirkwood and Riseman acknowledge that a polymer coil also has internal modes, in which the polymer's beads move with respect to each other.  These authors assume that the contribution of internal modes to the solution viscosity may be neglected.

Partial treatments of polymer internal modes were made by Rouse\cite{rouse1953a} and by Zimm\cite{zimm1956a}.  These authors approximate a polymer coil as a series of hydrodynamically active pointlike beads connected by hydrodynamically inert Hookian springs having unstretched length zero. The Rouse model does not have bead-bead hydrodynamic interactions.  In the Zimm model, bead-bead hydrodynamic interactions are described by the Oseen tensor. Rouse and Zimm each calculated the normal modes for a polymer coil containing $N$ beads, finding the $3N$ modes expected from classical mechanics.  However, they found three translational modes, and $3N-3$ orthogonal internal modes, modes in which the relative positions of beads depend on time.  Because all of the polymer's $3N$ modes were found to correspond to translation or to internal motions, Rouse and Zimm model polymers manifestly can not have rotational modes, modes in which the polymer center-of-mass is stationary, while the coil rotates without changing the distances between the beads.

How is this possible? As recently noted by this author\cite{phillies2018a}, the Rouse and Zimm models describe the internal modes of a polymer chain that is not subject to an applied shear. Rouse chains have no rotational modes because they are in an environment in which they are not called upon to rotate.  This author's molecular dynamics calculations\cite{phillies2018a} show that, when a fluid shear field is applied to a Rouse-model chain: (1) The polymer chain does rotate; (2) The Rouse modes become cross-correlated; and (3) The mean-square sizes and initial relaxation rates of the individual Rouse modes depend on the shear rate. In discussions of viscoelasticity, it is therefore more appropriate to refer to Rouse\emph{ coordinates} rather than to Rouse \emph{modes}. This is not to say that the Rouse and Zimm models are literally false.  However, these models only describe a polymer coil in a quiescent fluid, so they are irrelevant to discussions of polymer viscoelasticity.

The Kirkwood-Riseman and Rouse-Zimm models refer to the motions of a single isolated polymer chain.  In a series of papers, this author and collaborators extended the Kirkwood-Riseman model and the colloidal sphere calculations to treat hydrodynamic interactions between polymer chains\cite{phillies2016b,phillies1998a,phillies2002d,phillies1988c,phillies2002b,phillies1993b,merriam2004a}. In the model, bead motions are constrained; on the time scales of interest the net force and the net torque on each polymer coil must be very nearly zero.  The fluctuation-dissipation theorem ensures that the hydrodynamic interactions between diffusing polymer coils are the same as the hydrodynamic interactions between polymer coils, one of which is subject to an external force. No matter whether the chain motion is driven or diffusive, if a chain moves through solution it has an associated solvent flow, a wake in the nautical sense, that acts on other polymer chains, causing them to translate and rotate.  The motions of each of the other coils cannot be arranged in such a way that each bead of each coil is stationary with respect to the local solvent motion.  Instead, some of the beads must move with respect to the solvent, thereby exerting forces on the solvent and creating additional solvent flows.  Corresponding to these solvent flows and polymer motions, polymer transport coefficients become concentration-dependent.  On applying this description of hydrodynamic chain-chain interactions, mutually-consistent\cite{phillies2002d} power series for the concentration dependences of the polymer self-diffusion coefficient\cite{phillies1998a} and the viscosity increment due to the polymer\cite{phillies2002b} were obtained. All of the above calculations refer to a polymer as a curving line of spherical beads, much like the pearls on a necklace. Alternative descriptions of a polymer are possible.

More recently\cite{phillies1993b,merriam2004a}, the effects of higher-order hydrodynamics on polymer transport coefficients were considered. Considerations included the whole-chain/whole-chain hydrodynamic interaction and also the whole chain/single bead hydrodynamic interaction.  Higher order hydrodynamic interactions cause the diffusion and friction coefficients of a free monomer and of a monomer in a chain to depend on the polymer concentration in the solution. The concentration dependences are of the same nature but are not simply multiplicative, contrary to a common practice of correcting polymer solution transport data for so-called 'monomer friction effects' via normalization with respect to measurements of the friction coefficient of a free monomer.

With increasing concentration, polymer coils at first approach each other and then overlap. Overlap should lead to a dramatic increase in the importance of hydrodynamic interactions.  Why?  Consider polymer coils in a shear field.  As shown by Kirkwood and Riseman, the coils all rotate in the same direction, clockwise or counterclockwise. However, where two coils overlap, if the coils are both rotating in the same direction, clockwise or counterclockwise, then in the overlap region their beads are moving in opposite directions, creating a substantial additional torque on each chain. That torque arises because small numbers of beads on the two overlapping coils are close to each other, so that their hydrodynamic interactions become very important.  At the elevated concentrations at which the chains overlap, the net result of this additional torque is reasonably expected to be that whole-chain rotational motion is very greatly hindered relative to any hindrance of internal mode relaxations, leading at elevated concentrations to the appearance of a long-lived relaxational mode corresponding to chain reorientation. This paper represents a step toward calculating the contribution of hydrodynamic interactions to chain reorientation.

Our prior calculations of chain-chain hydrodynamic interactions did not include the one set of short-range hydrodynamic interactions that every polymer bead is sure to have, namely the short-range interactions between neighboring beads along the polymer coil.  A polymer chain has a correlation length, a distance over which the polymer approximates being straight, so one might envision an approximate polymer model in which the hydrodynamic units, modest pieces of a polymer coil, are linear groups of beads linked by free joints, rather than being single beads linked by freely-jointed springs.  As a physical image, in our possible model, a polymer would resemble the bar chains found attached to some ship's anchors, as opposed to resembling the pearls on a pearl necklace. Here we explore an aspect of this bar-chain model.

This note considers the effect of bead near-neighbors on bead-bead hydrodynamic interactions. We consider hydrodynamic properties of a single rod composed of beads in a line, between a bead and a rod, and between two such rods. We begin by calculating the mobility of a single rod. We then ask consider how the mobility of a bead or a  rod is affected by the presence of a second rod.

In the next Section of this note, analytic forms for short-range hydrodynamic interactions between freely-rotating spheres are presented. A further section shows how these interactions were used to calculate rod-rod hydrodynamic interactions. We then turn to our results on single-chain mobility, on bead-rod interactions, and on rod-rod interactions. A discussion closes the work.

\section{Underlying Theory}

We start by recalling the hydrodynamic interactions between colloidal spheres that are free to translate and rotate. Our objective is to determine the hydrodynamic interactions between assemblies of spheres, an assembly being a group of spheres that are mechanically linked to each other, as by Rouse-Zimm intermolecular springs, but that are not linked mechanically linked to any spheres other than the ones in the assembly. If it were desired to treat spheres, some of whose translations or rotations were obstructed, one could calculate the translational and rotational velocities being induced in the absence of the obstructions, and then add the additional forces and torques needed to block the translations and rotations.

Hydrodynamic interactions of freely rotating spheres were treated at length by Mazur and van Saarloos\cite{mazur1982a}.  Their calculation gives the velocity $\bm{u}_{i}$ induced in a sphere $i$ by the forces $\bm{F}_{j}$ exerted on the fluid by spheres $j$, in terms of the mobility $\bm{\mu}_{ij}$, written $\mu_{ij}^{\rm TT}$ in the original paper, namely
\begin{equation}
   \bm{u}_{i} = \frac{1}{6 \pi \eta a}\sum_{j} \bm{\mu}_{ij} \cdot \bm{F}_{j}.
         \label{eq:mudefinition}
\end{equation}

Here $\eta$ is the viscosity of the (assumed to be Newtonian) solvent; for simplicity, all spheres are taken to have the same radius $a$.   In the above sum $i=j$ is allowed.  While Mazur and van Saarloos did not use a simple method of reflections in their calculation, the spirit of that method is present in their description: A force applied on the solvent by a sphere $j$ propagates through a chain of zero or more other spheres $k, \ell, \ldots$ and eventually acts on the sphere of interest $i$, causing that sphere to move. We are considering creeping flows, for which inertia is negligible.  As a consequence, the net force and net torque on each sphere assembly must be very nearly zero, so that the hydrodynamic force and torque a sphere assembly exerts on the solvent must very nearly equal to the external force and torque on that assembly.  Mazur and van Saarloos also consider the translational velocity induced in $i$ if $j$ applies a torque to the surrounding fluid, and the rotational velocity $\bm{\omega}_{i}$ induced in sphere $i$ by a force or torque applied to the fluid by sphere $j$.  Their calculation as summarized here begins with eq. \ref{eq:mudefinition}.

The matrix $\bm{\mu}_{ij}$ as developed by Mazur and van Saarloos is written as a sum of one-, two-, three-, and four-particle terms, where the same particle may appear more than once in the sum.  Terms involving more than four particles exist, but were not obtained in their development. The one-particle term is
\begin{equation}
   \bm{\mu}_{ij1} = \delta_{ij} \bm{I},
\label{eq:mu1}
\end{equation}
where $\bm{I}$ is the unit identity matrix.

For the two-particle terms with $i \neq j$, they note the standard
\begin{equation}
    \bm{\mu}_{ij2} = \frac{3a}{4r_{ij}}(\bm{I} + \hat{r}_{ij}\hat{r}_{ij}) - \frac{a^{3}}{2 r_{ij}^{3}}(3 \hat{r}_{ij}\hat{r}_{ij} - \bm{I}).
    \label{eq:mu2}
\end{equation}
Here  $r_{ij}$ is the distance between spheres $i$ and $j$, and $\hat{r}_{ij}$ is the unit vector directed from $i$ toward $j$.  Terms of $\bm{\mu}_{ij2}$, in which $i = j$, vanish.

The three-particle term is
\begin{displaymath}
  \bm{\mu}_{ikj} =  - \left(\frac{15}{8}\right)\left(\frac{a^{4}}{r_{ik}^{2} r_{kj}^{2}}\right)\left( 1 - 3(\hat{r}_{ik} \cdot \hat{r}_{kj})^{2} \right)\hat{r}_{ik} \hat{r}_{kj}
  \end{displaymath}
\begin{displaymath}
    +\left( \frac{3 a^{6}}{r_{ik}^{2} r_{kj}^{4}}\right) \left[\left(1-5(\hat{r}_{ik} \cdot \hat{r}_{kj})^{2}\right)\hat{r}_{ik}\hat{r}_{kj} + 2(\hat{r}_{ik} \cdot \hat{r}_{kj}) \hat{r}_{ik}\hat{r}_{ik} \right]
\end{displaymath}
\begin{displaymath}
    + \left( \frac{3 a^{6}}{r_{ik}^{4} r_{kj}^{2}}\right) \left[\left(1-5(\hat{r}_{ik} \cdot \hat{r}_{kj})^{2}\right)\hat{r}_{ik}\hat{r}_{kj} + 2(\hat{r}_{ik} \cdot \hat{r}_{kj}) \hat{r}_{kj}\hat{r}_{kj} \right]
\end{displaymath}
\begin{displaymath}
     + \frac{64 a^{6}}{r_{ik}^{3}r_{kj}^{3}} \left[\left(49-117  (\hat{r}_{ik} \cdot \hat{r}_{kj})^{2}\right) \bm{I} \right.
\end{displaymath}
\begin{displaymath}
      -\left(93-315(\hat{r}_{ik} \cdot \hat{r}_{kj})^{2}\right)\left( \hat{r}_{ik}\hat{r}_{ik} + \hat{r}_{kj}\hat{r}_{kj}\right) + 54 (\hat{r}_{ik} \cdot \hat{r}_{kj}) \hat{r}_{kj} \hat{r}_{ik})
\end{displaymath}
\begin{equation}
    +  \left. \left(729 - 1575(\hat{r}_{ik} \cdot \hat{r}_{kj})^{2}  \right) (\hat{r}_{ik} \cdot \hat{r}_{kj})\hat{r}_{ik} \hat{r}_{kj} \right].
   \label{mu3}
\end{equation}
The constraints on the indices are $i \neq k$ and $k \neq j$.  Terms of $\bm{\mu}_{ikj}$ in which its three indices do not satisfy these two constraints vanish. Terms in which $i=j$ are allowed.

The four-particle term is
\begin{displaymath}
  \bm{\mu}_{iklj} = \left(\frac{75}{16}\right) \left( \frac{a^{7}}{r_{ik}^{2} r_{k\ell}^{3} r_{\ell j}^{2}} \right) \left[\left(1 - 3 ( \hat{r}_{ik} \cdot \hat{r}_{k\ell})^{2} \right) \left(1-3 (\hat{r}_{k\ell} \cdot \hat{r}_{\ell j})^{2} \right)      \right.
\end{displaymath}
\begin{equation}
     \left. + 6( \hat{r}_{ik} \cdot \hat{r}_{k\ell})^{2} (\hat{r}_{k\ell} \cdot \hat{r}_{\ell j})^{2} - 6 (\hat{r}_{ik} \cdot \hat{r}_{k \ell})  (\hat{r}_{k\ell} \cdot \hat{r}_{\ell j})(\hat{r}_{\ell j} \cdot \hat{r}_{ik})\right] \hat{r}_{ik} \hat{r}_{\ell j}
\label{eq:mu4}
\end{equation}
with the constraints $i \neq k$, $k \neq \ell$, and $\ell \neq j$. Terms of $ \bm{\mu}_{iklj}$ in which its four indices do not satisfy these constraints vanish.

In the above, $\bm{\mu}_{ij2}$ may be recognized as the Oseen and Rotne-Prager tensors. The above terms combine to give a $3\times3$ mobility tensor, namely
\begin{equation}
    \bm{\mu}_{ij} = \delta_{ij} \bm{I} + \bm{\mu}_{ij2} +\sum_{k} \bm{\mu}_{ikj} +\sum_{k} \sum_{\ell} \bm{\mu}_{ik\ell j}.
\label{eq:mutotal}
\end{equation}
The limits on the sums are discussed below.

\section{Extension to Assemblies of Spheres}

We now translate the Mazur-van Saarloos discussion, which was a consideration of spheres each free to move with respect to the other spheres, to a consideration of rigid assemblies of spheres. To link to polymer theory, the individual spheres are henceforth referred to as 'beads', as seen in the polymer bead-spring models. Our interest here is the interaction of rigid multibead assemblies. In the detailed calculations of the following sections, all assemblies are nearly linear rods, one bead wide and one or $N$ beads long.

A simple generalization of the previous Section's notation permits us to advance in a clear manner.  The components of the forces that the first cluster in the calculation exerts on the solvent are collected into a $3N$-dimensional vector $\widetilde{\bm{F}}$. The first three components of  $\widetilde{\bm{F}}$ are the three components of the force $\bm{F}_{1}$ exerted on the solvent by particle $1$, the next three components of
$\widetilde{\bm{F}}$ are the three components of the force $\bm{F}_{2}$ exerted on the solvent by particle $2$, and so forth.  The velocities of the $N$ beads in a rod may similarly be written as a single $3N$-dimensional vector $\widetilde{\bm{U}}$, namely
\begin{equation}
  \widetilde{\bm{U}}  = \{\bm{u}_{1}, \bm{u}_{2}, \ldots, \bm{u}_{N} \},
\label{eq:Udefinition}
\end{equation}
with $\bm{u}_{1}$ being the velocity induced in the first bead, etc.

$\widetilde{\bm{F}}$ and $\widetilde{\bm{U}}$ are related by a generalization of eq.\ \ref{eq:mudefinition}, namely
\begin{equation}
     \widetilde{\bm{U}} =   \widetilde{\bm{\mu}} \cdot \widetilde{\bm{F}}
\label{eq:Nparticlemobility}
\end{equation}
in which $\widetilde{\bm{\mu}}$ is the $3N \times 3N$ mobility tensor linking the forces to the induced velocities. For simplicity we chosoe units such that $6 \pi \eta a = 1$. The matrix $\widetilde{\bm{\mu}}$ is assembled from the 3x3 matrices given in the previous section, so that $\bm{\mu}_{11}$ occupies the upper left corner of  $\widetilde{\bm{\mu}}$, while $\bm{\mu}_{12}$, $\bm{\mu}_{13}$, \ldots fill the remaining columns and the first three rows, $\bm{\mu}_{22}$, $\bm{\mu}_{33}$, etc., advance down the main diagonal of  $\widetilde{\bm{\mu}}$, one $3 \times 3$ square block at a time, and so forth.

In the historical work\cite{bloomfield1967a,bloomfield1967b,phillies1984z}, the mobility of a single assembly of $N$ beads was considered. The indices $i$ and $j$ (and, as needed, $k$ and $\ell$) each ranged over the $N$ beads. Equation \ref{eq:Nparticlemobility} was then solved to calculate the forces needed to drive the assembly with unit velocity along each of the Cartesian axes. To do this, the $3N$-dimensional velocity vector was taken to be $\widetilde{\bm{U}_{x}} = {1,0,0,1,0,0,\ldots,1,0,0}$ for unit velocity of the rigid assembly in the $x$-direction, and similarly for unit velocities in the $y$ or $z$ directions. On evaluating numerically all the components of $\widetilde{\bm{\mu}}$, eq.\ \ref{eq:Nparticlemobility} becomes a set of $3N$ linear equations in $3N$ unknowns, the unknowns being the $3N$ components of $\widetilde{\bm{F}}$. Standard linear algebraic methods give a solution for $\widetilde{\bm{F}}$, these being the forces needed to cause all beads to move with unit velocity in the $x$-direction. In general, the forces on different beads are not equal to each other. Because the assembly is moving as a rigid body, internal stresses associated with bead-bead bonds must sum to zero.

One then calculates the total force $\vec{F}^{T}$ needed to move the assembly parallel to Cartesian coordinate $n$, $n \in (1,3)$. The components $F^{T}_{n}$ of this force are obtained from the components of $\widetilde{\bm{F}}$ as
\begin{equation}
    F^{T}_{n} = \sum_{j=1}^{N} \widetilde{\bm{F}_{3(j-1)+n}}.
\label{eq:totalforce}
\end{equation}
For the simple arrays of spheres considered here, $\vec{F}^{T}$ is simply parallel to $\widetilde{\bm{U}}$, so correspondingly the drag tensor is effectively a set of three scalars $\zeta_{n}$, one for each direction. As $F^{T}_{i}$ gives the force needed to produce a unit velocity of the assembly, the inverse
\begin{equation}
   \zeta_{n} = (F^{T}_{n})^{-1}
   \label{eq:dragcoefficients}
\end{equation}
gives the velocity produced by a unit force.  Furthermore,
\begin{equation}
     U_{n}  = \zeta_{n}  F_{n}
\label{eq:Uassembly}
\end{equation}
gives the velocity in the $n$-direction as created by a force $F_{n}$ in the $n$ direction. The $\zeta_{n}$ are the mobilities of the assembly, if the assembly is treated as a single object.

For the calculations here, the assemblies are linear rods, represented as lines of beads. The beads have unit radius, i.e., $a=1.0$.  Adjoining beads are separated by $2.5 a$. To eliminate the zero eigenvalue associated with rotation of each rod around its long axis, bead locations were cyclically perturbed perpendicular to their rod's axis by $0.001a$. To assess the effects of rod-rod interactions, we did additional calculations after adding to the model a second rod of beads. The first assembly was then either a single bead or a rod having $N$ beads. The second rod always contained $N$ beads. The first rod was taken to lie on the $x$ axis.  The second rod was displaced from the first in the $+y$ direction; the centers of the two rods share the same $x$ and $z$ coordinates.  In different calculations, the axis of the second rod was taken to be parallel to the $x$ axis ('parallel' configuration) or parallel to the $z$ axis ('perpendicular' configuration).

The two rods were aligned symmetrically so that the motion of one rod would create no net torque on the second rod, and hence would create no net tendency of the second rod to rotate. From symmetry, for these configurations the mobility tensor reduces to its three diagonal elements, namely $\zeta_{1}$, $\zeta_{2}$, and $\zeta_{3}$.  If the motion of one rod drove the other to reorient itself, it would be necessary to add to the calculation the translation-rotation, rotation-translation, and rotation-rotation coupling tensors treated by Mazur and van Saarloos\cite{mazur1982a}.  We did not do this here.

As a separate issue, we here assumed that the individual beads are free to rotate. However, in a rod, each bead is connected to one or two neighbors.  Those connections might interfere with rotation, hypothetically creating torques on the individual beads that would create additional hydrodynamic couplings within and between bead assemblies.

In numerical calculations using the equations from the previous Section, the sums over the indices $k$ and $\ell$ proceed over all beads in the system. The sums over $i$ proceed over the beads or bead of the first rod.  For the sums over $j$, there are two alternatives.  To evaluate the self-mobility of the first rod, the sum on $j$ proceeds over all beads of the first rod.  To evaluate the cross-mobility, the effect of the first rod's motion on the motions of a second rod, the sum on $j$ would need to proceed over all beads of the second rod. We did not study the cross mobility here.  Preliminary calculations were done in Fortran.  The calculation was then converted to Mathematica to take advantage of Mathematica's extreme-numerical-precision features.

\section{Results for Self-Mobility}

\begin{figure}[htb]
  \centering
  \includegraphics[width=4.5in]{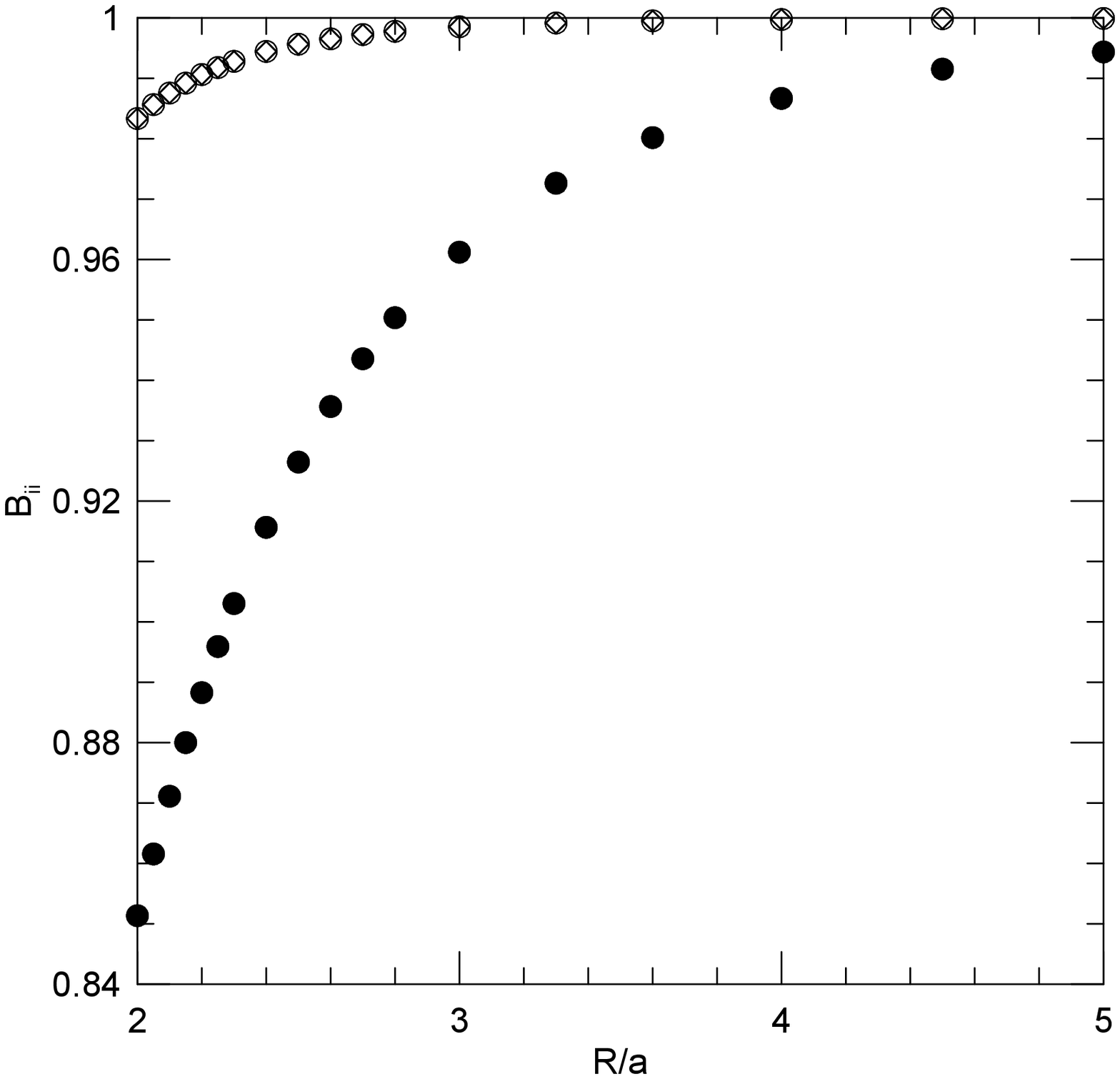}
  \caption{Mobility $\bm{B}_{ii}$ of a sphere having unit radius, in the presence of a second sphere also having unit radius, as a function of the distance between their centers. $R/a = 2$ is the distance between the centers of the two spheres when the spheres are in contact. $\bm{B}_{ii}$ is anisotropic, with $B_{yy}$ (filled circles) for motion in the $y$ direction, parallel to the line of centers, depending on $R/a$ much more strongly than does $B_{xx}$ (open circles) or $B_{zz}$ (diamonds, superposed on circles) for motion in the $x$ or $z$ directions, perpendicular to the line of centers. }\label{figure1}
\end{figure}

We now consider results for the self-mobility tensor $\bm{B}$. The self-mobility of the first rod is described by a $3 \times 3$ matrix $\bm{B}_{ij}$. For the simple configurations considered here, $\bm{B}_{ij}$ is diagonal with $B_{nn} = \zeta_{n}$.  Figure \ref{figure1} shows the diagonal components $B_{ii}$ of this tensor, for a pair of beads separated along the $y$ axis by $R$. For this system, by symmetry $B_{xx} = B_{zz}$. The effect of hydrodynamic interactions on $\bm{B}$ of the first bead is anisotropic, with motion parallel to the line of centers of the two beads being resisted more than is motion perpendicular to the line of centers. The dependence of $\bm{B}$ on $R$ only appears because we have included hydrodynamic interactions through to the $(a/R)^{7}$ level.  If we had approximated hydrodynamic interactions at the level of the Oseen or Rotne-Prager tensors, our calculations would instead have found that $\bm{B}$ is independent of $R$.

[Figure One Here]

\begin{figure}[htb]
  \centering
  \includegraphics[width=5in]{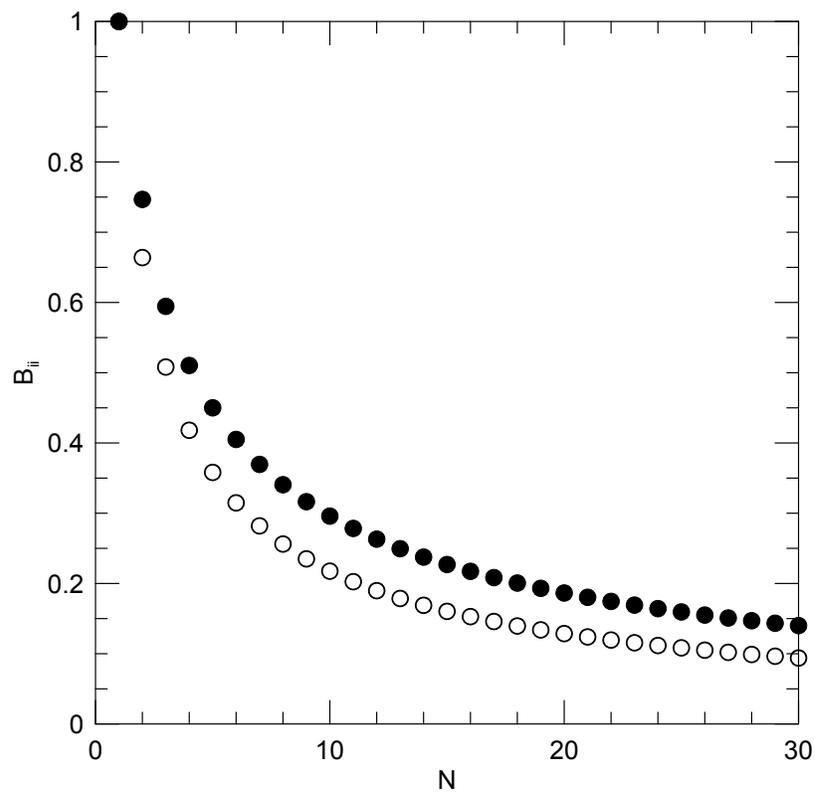}
  \caption{Mobility of a single rod of $N$ beads as a function of the length $N$ of the rod, including  mobility parallel to the line of centers (filled points) and mobility perpendicular to the line of centers (open points).} \label{figure2}
\end{figure}

We now apply the above hydrodynamic treatment to calculate the mobility of a single rod of $N$ beads as a function of the length $N$ of the rod. Results appear in Figure \ref{figure2}. The mobility of a rod parallel to its long axis is appreciably  greater than the same rod's mobility perpendicular to its long axis.  For a 20-bead rod, $B$ for parallel motion is 45\% greater than is $B$ for motion perpendicular to the rod's line of centers.  Hydrodynamic interactions also serve to shield the interactions of individual beads from the solvent.  Both mobilities decrease with increasing $N$ more slowly than $1/N$. The 20-bead rod, instead of having 1/20th the mobility of a single bead, has slightly less than 1/5th the mobility of a single bead for motion parallel to its long axis, and slightly more than 1/8th the mobility of a single bead for motion perpendicular to its long axis.

[Figure Two Here]

\begin{figure}[htb]
  \centering
  \includegraphics[width=5in]{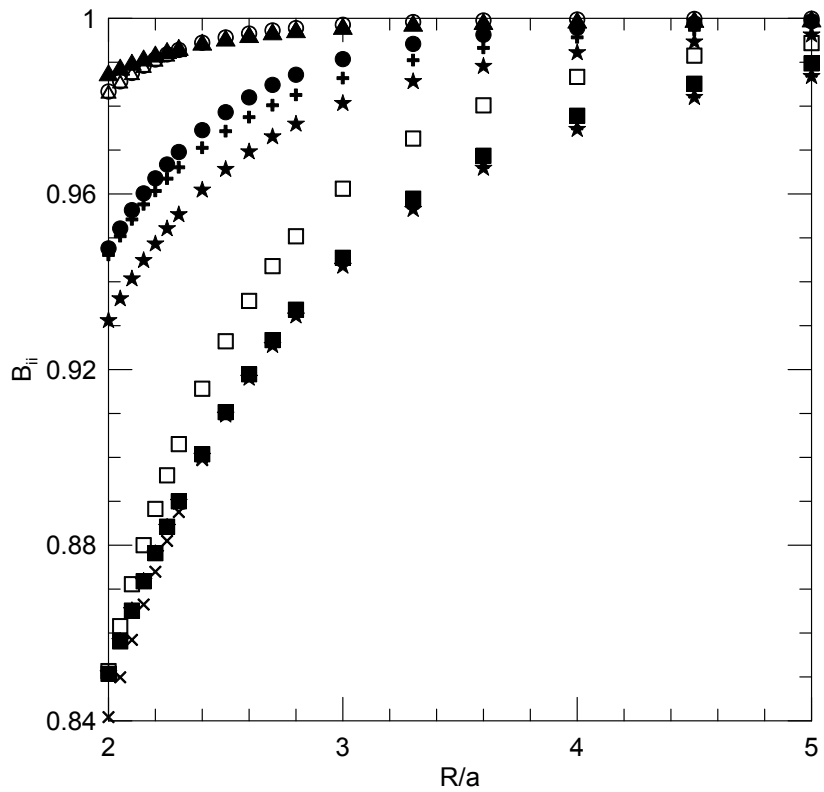}
  \caption{Mobility $B_{ii}$ of a single bead in the vicinity of a single bead or a 2-bead, 3-bead, or 10-bead rod, as a function of the distance between the single bead and the line of centers of the rod. The rod lies parallel to the $x$-axis.  The bead and rod are separated along the $y$-axis. The points at the top of the graph ($\blacktriangle$ and neighboring points) represent $B_{xx}$ and $B_{zz}$ for two beads, and $B_{zz}$ for a single bead moving near a rod of any length. $B_{xx}$ for a bead and a two-bead ($\bullet$), three-bead ($+$), or ten-bead ($\bigstar$) rod lie lower in the graph, with $B_{xx} \approx 0.95$ at $R/a = 2$.  Squares mark $B_{yy}$ (motion of the bead directly toward or away from the rod) for a pair of beads ($\square$) or a bead and a two-bead rod ($\blacksquare$). Points for a bead moving toward or away from a 3-bead ($X$) or a 10-bead ($\bigstar$) rod are largely masked by the points for a bead and a two-bead rod.}\label{figure3}
\end{figure}

Figure \ref{figure3} shows the three components of $\bm{B}$ for a single bead in the presence of another bead or a rod containing two, three, or ten beads, as a function of the distance $R/a$ between the first bead and the rod's center. At short distances $R/a \leq 4.0$ or so, $B_{xx}$ and $B_{yy}$ depend markedly on $R/a$, while $B_{zz}$ is nearly constant.
As seen on the graph, $B_{xx}$, describing motion of the bead parallel to the adjoining rod's line of centers, decreases visibly with increasing rod length.  $B_{yy}$ for a bead near a rod is 1-2\% less than $B_{yy}$ for two nearby beads, but is very nearly independent of how long the rod is. $B_{zz}$ for the single bead is nearly independent of the length of the neighboring rod, and is always $> 0.98$.

[Figure Three Here]

\begin{figure}[htb]
  \centering
  \includegraphics[width=5in]{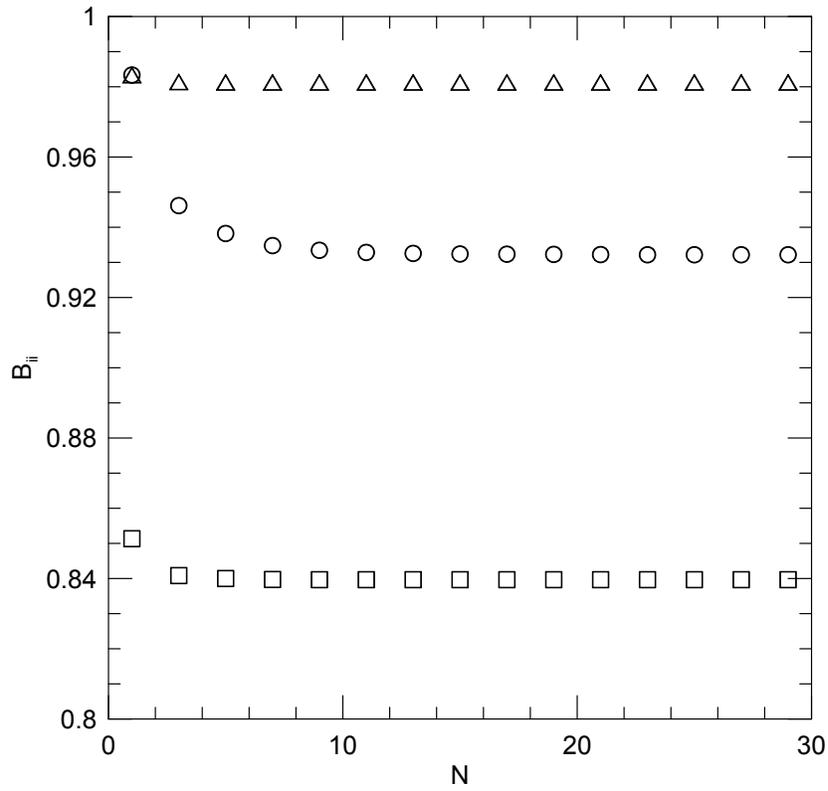}
\caption{Mobilities $B_{xx}$ ($\circ$), $B_{yy}$ ($\square$), and $B_{zz}$ ($\vartriangle$) of a single bead in contact ($R/a = 2$) with the center bead of a rod of length $N$, as functions of $N$. Because the rod is chosen to lie symmetrically with respect to the bead, and the single bead and the center bead of the rod are in contact, $N$ must be an odd number.}\label{figure4}
\end{figure}

Figure \ref{figure4} shows the three components of the mobility $B_{ii}$ of a single bead that is in contact with the center of a rod having length $N$, as a function of $N$.  The $B_{ii}$ only depend perceptibly on $N$ for $N \leq 9$. We see $B_{yy} < B_{xx} < B_{zz}$.   $B_{xx}$ falls from 0.98 to 0.93 as the length of the second rod is increased from one bead to nine beads. $B_{yy}$ and $B_{zz}$ are, except for the very shortest rods, nearly independent of $N$.   For $B_{yy}$ or $B_{zz}$, but not $B_{xx}$, the single bead-single bead interaction form is a good approximation to the bead-rod interaction.  As revealed by $B_{xx}$, a bead finds it more difficult to slide along the length of a long rod than it does to move in either of the directions perpendicular to the rod's long axis.

[Figure Four Here]

\begin{figure}[htb]
  \centering
  \includegraphics[width=5in]{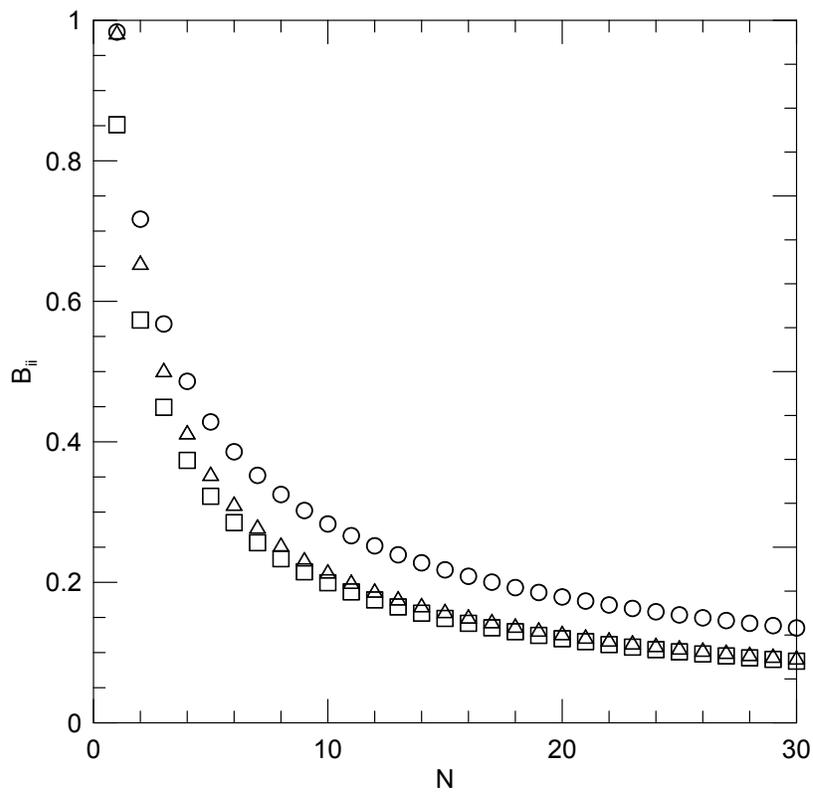}
  \caption{Mobilities $B_{xx}$ ($\circ$), $B_{yy}$ ($\square$), and $B_{zz}$ ($\vartriangle$) of two parallel rods having equal lengths $N$, as functions of $N$. The rods lie parallel to the $x$-axis and are in contact with each other.} \label{figure5}
\end{figure}

\begin{figure}[htb]
  \centering
  \includegraphics[width=5in]{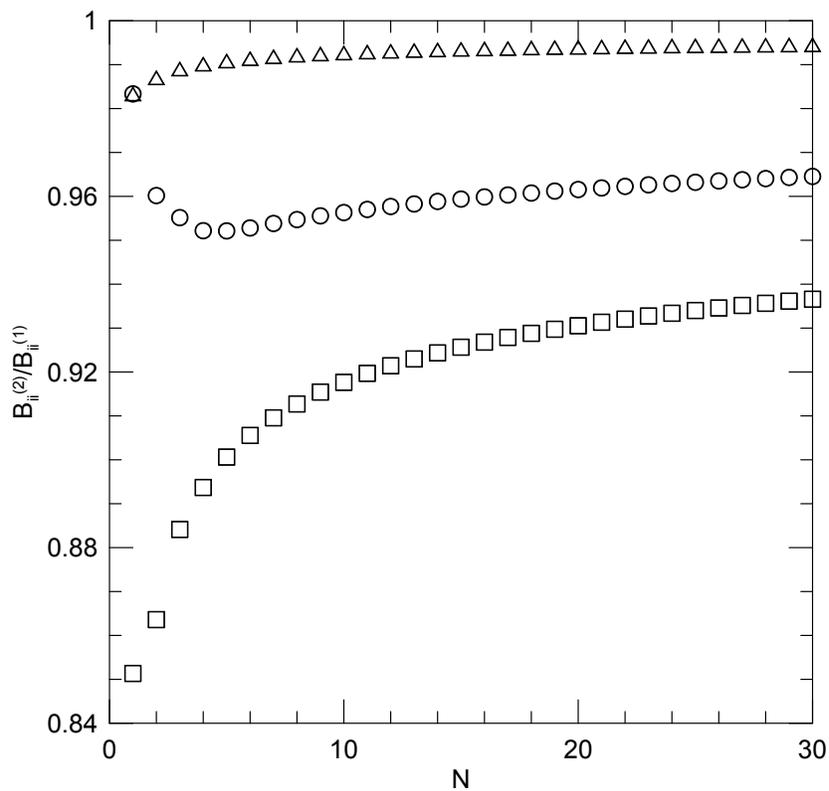}
  \caption{Mobilities $B_{xx}^{(2)}$ ($\circ$), $B_{yy}^{(2)}$ ($\square$), and $B_{zz}^{(2)}$ ($\vartriangle$) of two parallel rods having equal lengths $N$, as functions of $N$.  The mobilities were normalized, respectively, by the mobilities $B_{xx}^{(1)}$, $B_{yy}^{(1)}$, and $B_{zz}^{(1)}$  of isolated rods having the same lengths $N$, thereby revealing the effect of the second rod on the mobility of the first rod. The rods lie parallel to the $x$-axis and are in contact with each other.} \label{figure6}
\end{figure}

Figures \ref{figure5} and \ref{figure6} show the effects of rod-rod interactions for two rods of length $N$ lying parallel to each other along the $x$ axis, as a function of the length of the rods.  The two rods are separated by $2.0a$, so their neighboring beads are in contact. From Figure 5, $B_{xx}$ is consistently larger than $B_{yy}$ or $B_{zz}$.  For shorter rods, $N < 10$, $B_{yy} < B_{zz}$.  For longer rods, i.e., $N > 10$, $B_{yy} \approx B_{zz}$.  Figure \ref{figure6} shows the extent to which the presence of the second rod affects the motion of the first.  We plot the ratio of $B_{ii}^{(2)}$, the mobility of the first rod when both rods are present, to $B_{ii}^{(1)}$, the mobility of the first rod when only one rod is present, as a function of rod length.  For $B_{xx}$ and $B_{zz}$, the effect of the second rod is nearly independent of rod length, reducing $B_{ii}$ by approximately 4\% and 1\%, respectively, from its isolated-single-rod value.  For $B_{yy}$, the effect of the second rod depends strongly on rod length, the effect of the second rod decreasing as the rod length is increased.  The reduction in $B_{yy}$ by the second rod is by 16\% for a pair of beads, by less than 9\% for a pair of $N=10$ rods, and then tends toward a possible asymptotic limit (we did not reach the asymptote, which may not exist) giving a reduction in $B_{yy}$ due to the second rod of 6\% or so at large $N$.

[Figures Five and Six Here]

Figures 7 and 8 show the dependence of $B_{ii}$ on distance for pairs of short rods.  We contrast $B_{ii}$ for two spheres, for two parallel three-bead rods, and for two parallel five-bead rods, viewed as functions of the distance $R/a$ between the rods.  Figure 7 shows actual values of $B_{xx}$, $B_{yy}$, and $B_{zz}$ as functions of $R/a$.  The $B_{ii}$ decrease monotonically with increasing rod length. Except at short distances $R/a < 3.0$, $B_{ii}$ for each rod pair and direction of motion is nearly independent of $R/a$. At shorter distances $R/a < 3.0$, $B_{yy}$ and to a lesser extent $B_{xx}$ decrease with decreasing $R/a$.

Figure 8 shows the effect of the second rod on the mobility of the first rod, plotting  $B_{ii}^{(2)}/B_{ii}^{(1)}$ against $R/a$ for all three rod lengths.  The effect of the second rod is smallest for $B_{zz}$, so that $B_{zz}^{(2)}/B_{zz}^{(1)}$ is $> 0.98$ at all distances for all rod pairs.  The effect of the second rod is largest for $B_{yy}$, so that $B_{yy}^{(2)}/B_{yy}^{(1)}$  at small $R/a$ is as small as 0.85-0.9. However, $B_{yy}^{(2)}/B_{yy}^{(1)}$ depends less strongly on rod length than it depends on $R/a$, so that the three $B_{yy}^{(2)}/B_{yy}^{(1)}$ curves stay fairly close together for $R/a > 2.2$.  The behavior of $B_{xx}^{(2)}/B_{xx}^{(1)}$ is intermediate in several respects.  In particular, $B_{xx}^{(2)}/B_{xx}^{(1)}$  changes little between the 3-bead and the 5-bead rods, but depends on $R/a$  more strongly for the rods than for a pair of beads. At $R/a=2$, for a pair of beads $B_{xx}^{(2)}/B_{xx}^{(1)} \approx 0.99$, while for rods at the same separation $B_{xx}^{(2)}/B_{xx}^{(1)} \approx 0.95$.

[Figures Seven and Eight Here]

Instead of being parallel to each other, the two rods could be perpendicular.  We take the first rod to lie along the $x$-axis, and the second to lie parallel to the $z$-axis. The centers of the two rods lie on the $y$-axis.  Figure \ref{figure9} shows the $x$, $y$, and $z$ components of the normalized mobilities $B_{ii}^{(2)}/B_{ii}^{(1)}$ for two perpendicular rods of length $N$ whose midpoints are separated by $2.0 a$. Except for the very shortest rods, the second rod has almost not effect on the mobility of the first, because most beads of each rod are too far apart to interact with each other via the $(a/R)^{4}$ interaction. The small oscillation between even and odd $N$ arises because for odd $N$ the two beads at the rod's centers are in contact, while for even $N$ the nearest contact between the two rods is at the midpoints between two pairs of beads.

[Figure Nine Here]

\section{Discussion}

In this note we considered the effect of hydrodynamic interactions on the mobility of an isolated rod, and the effect of hydrodynamic interactions on the mobility of a single bead or a rod, due to  a second rod placed parallel or perpendicular to the first.

Hydrodynamic interactions have a large effect on the mobility of a single rod of beads.  As is well-known, the mobility of a long rod parallel to its own axis is half again as large as its mobility perpendicular to its long axis.  Hydrodynamic interactions also reduce the contributions to a rod's frictional resistance of the individual beads in a rod.  The mobility of a 30-bead rod is three or five times larger than would be the mobility of 30 beads that lacked hydrodynamic interactions with each other.

The mobility of an isolated bead near a long rod is largely independent of the rod's length, even when the bead and rod are almost in contact, and is not very different from the mobility of an isolated bead in the presence of a second single bead. The effect of hydrodynamic interactions on the mobility of a bead, due to the presence of a second bead or a rod, falls off with the distance between the first bead and the rod.  However, even for short rods the distance dependence depends only weakly on how long the second rod is.

We also considered the hydrodynamic interaction between two parallel rods of equal length.  The mobility of an isolated rod depends strongly on the rod's length.  The longer a rod is, the smaller its mobility is. Juxtaposing a second rod to the first reduces the first rod's mobility, though not by a great deal.  However, the reduction in the first rod's mobility by the presence of the second rod depends only weakly (cf.\ Figure 8) on the length of the two rods. In addition, the mobility of a given rod is reduced only slightly (cf.\ Figure 9) by the presence of a second perpendicular rod, even when the two rods are in near-contact at their centers.

Our results  are understandable in terms of the short-range nature of parts of the hydrodynamic interaction tensors.  If we had performed our calculations at the Oseen or Rotne-Prager tensors, the presence of a second bead or rod would have no effect on the mobility of the first bead or rod.  On the same line, two perpendicular rods have little effect on each other's mobility, because most beads of each rod are too far away from the beads of the other rod to interact with them appreciably.

The large difference in $B_{xx}$ between single bead-single bead and single bead-short rod systems arises from the tensor nature of the hydrodynamic interactions and the extended nature of the second rod.  Consider a single bead moving parallel to a long rod.  The vector from the single bead to the nearest bead of the rod is almost perpendicular to the first bead's direction of motion, so the contribution of the nearest bead to reducing the single bead's mobility is small.  However, the vector from the single bead to the other beads in the rod includes components both perpendicular to and parallel to the single bead's velocity.  This parallel component makes a larger contribution to the bead's $B_{xx}$ than the nearest bead did, so as a qualitative result $B_{xx}$ for a bead and a long rod is appreciably smaller than $B_{xx}$ for two beads.

These calculations lead to a central conclusion. The hydrodynamic interactions between two rods, as they affect the mobility of a given rod, are not very different from the interactions between two beads, especially when the two rods are even modestly separated, so at the level of approximation of bead-spring models of polymer coils it does not appear to be advantageous to replace sphere-sphere hydrodynamic interaction tensors with more elaborate rod-rod hydrodynamic interaction tensors.

\pagebreak

\pagebreak

\pagebreak

\pagebreak

\pagebreak

\pagebreak

\pagebreak

\begin{figure}[htb]
  \centering
  \includegraphics[width=5in]{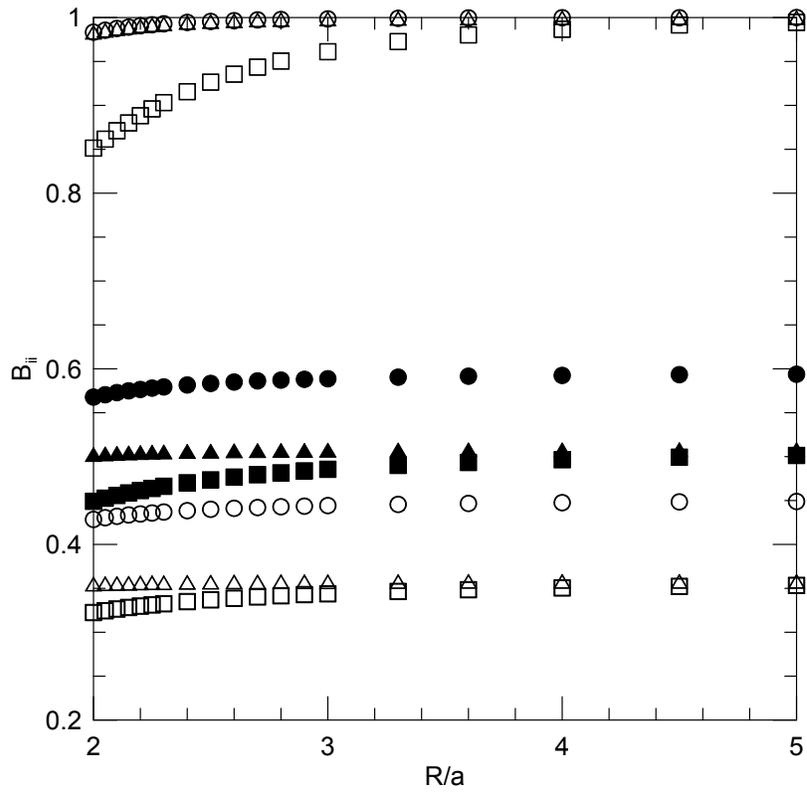}
  \caption{Mobilities $B_{xx}$ ($\circ$), $B_{yy}$ ($\square$), and $B_{zz}$ ($\vartriangle$) of two parallel rods having equal lengths $N$, as functions of $R/a$. The rods are parallel to the $x$-axis and have lengths $N=1$ (open points, top of figure), $N=3$ (filled points), and $N=5$ (open points, bottom of figure).} \label{figure7}
\end{figure}

\pagebreak

\begin{figure}[htb]
  \centering
  \includegraphics[width=5in]{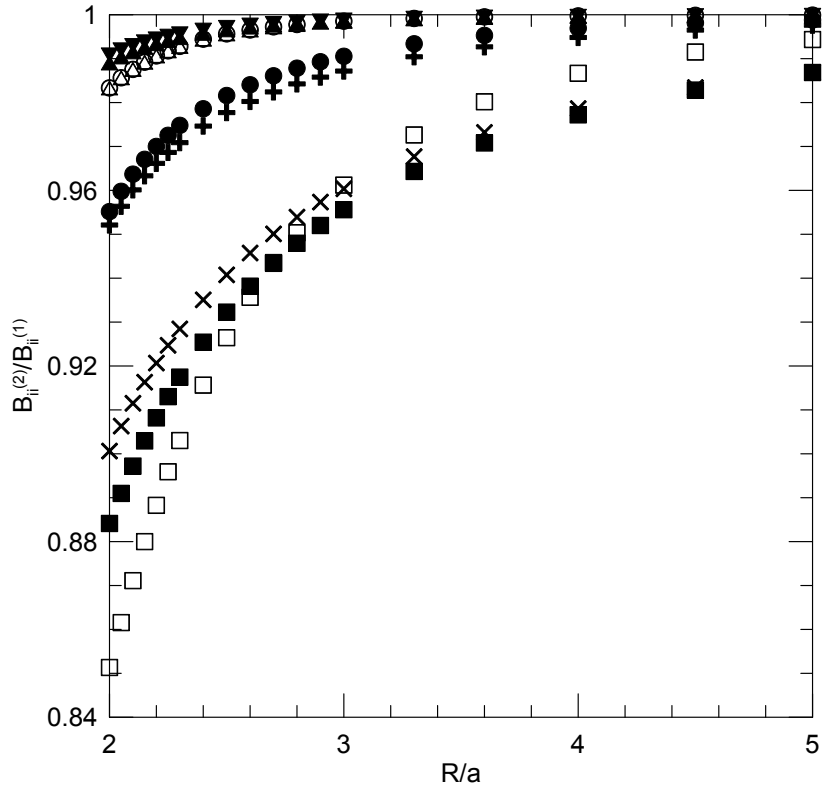}
  \caption{Mobilities $B_{xx}^{(2)}$, $B_{yy}^{(2)}$, and $B_{zz}^{(2)}$ of two parallel rods having equal lengths $N$, as normalized by the corresponding mobilities $B_{xx}^{(1)}$, $B_{yy}^{(1)}$, and $B_{zz}^{(1)}$ of an isolated rod having the same length, as functions of $R/a$. The rods are both parallel to the $x$-axis.  Point shapes, in the order ($B_{xx}$, $B_{yy}$, $B_{zz}$) for each rod length, are ($\circ$, $\square$, $\vartriangle$) for $N=1$,  ($\bullet$, $\blacksquare$, $\blacktriangle$) for $N=3$, and ($+$, $\times$, $\blacktriangledown$) for N=5.} \label{figure8}
\end{figure}

\begin{figure}[htb]
  \centering
  \includegraphics[width=5in]{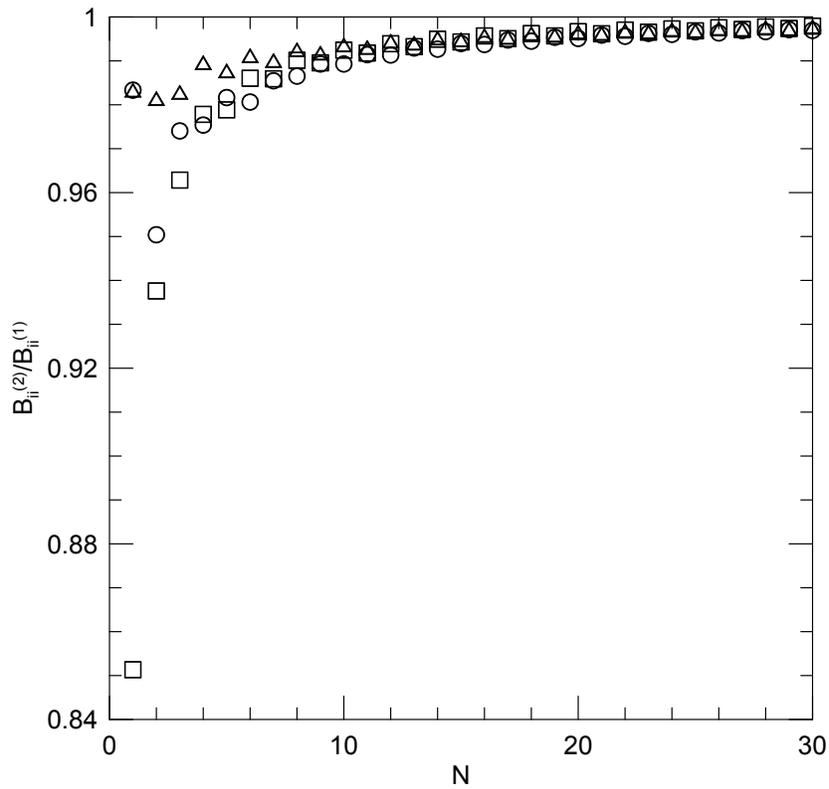}
  \caption{Mobilities $B_{xx}$ ($\circ$), $B_{yy}$ ($\square$), and $B_{zz}$ ($\vartriangle$) as functions of rod length, for two perpendicular rods lying along the $x$ and $z$ axes, the two rods being in closest contact at their midpoints, with their lines of centers being separated by $2.0a$.  Small oscillations appear because at point of closest contact between the two rods there may be two beads, or there may be the midpoints between two pairs of beads.} \label{figure9}
\end{figure}

\end{document}